\documentclass[journal,comsoc]{IEEEtran}

\usepackage[T1]{fontenc}

\usepackage{ifpdf}
\usepackage{cite}
\usepackage{graphicx}
\graphicspath{{./Fig/}}
\usepackage{amsmath,amsfonts}
\usepackage{bm}
\usepackage{algorithm}
\usepackage{algorithmic}
\usepackage{array}
\usepackage[caption=false,font=normalsize,labelfont=sf,textfont=sf]{subfig}
\usepackage{url}
\begin{document}

\title{Development of Advanced FEM Simulation Technology for Pre-Operative Surgical Planning}

\author{Zhanyue Zhao, Yiwei Jiang, Charles Bales, Yang Wang, Gregory Fischer
\thanks{Zhanyue Zhao, Yiwei Jiang, Charles Bales, Yang Wang, and Gregory Fischer are with the Department of Robotics Engineering, Worcester Polytechnic Institute, Worcester, MA 01605 USA (e-mail: zzhao4@wpi.edu, gfischer@wpi.edu).}
\thanks{This research is supported by National Institute of Health (NIH) under the National Cancer Institute (NCI) under Grant R01CA166379 and R01EB030539.}
}
\maketitle

\begin{abstract}

Intracorporeal needle-based therapeutic ultrasound (NBTU) offers a minimally invasive approach for the thermal ablation of malignant brain tumors, including both primary and metastatic cancers. NBTU utilizes a high-frequency alternating electric field to excite a piezoelectric transducer, generating acoustic waves that cause localized heating and tumor cell ablation, and it provides a more precise ablation by delivering lower acoustic power doses directly to targeted tumors while sparing surrounding healthy tissue. Building on our previous work, this study introduces a database for optimizing pre-operative surgical planning by simulating ablation effects in varied tissue environments and develops an extended simulation model incorporating various tumor types and sizes to evaluate thermal damage under trans-tissue conditions. A comprehensive database is created from these simulations, detailing critical parameters such as CEM43 isodose maps, temperature changes, thermal dose areas, and maximum ablation distances for four directional probes. This database serves as a valuable resource for future studies, aiding in complex trajectory planning and parameter optimization for NBTU procedures. Moreover, a novel probe selection method is proposed to enhance pre-surgical planning, providing a strategic approach to selecting probes that maximize therapeutic efficiency and minimize ablation time. By avoiding unnecessary thermal propagation and optimizing probe angles, this method has the potential to improve patient outcomes and streamline surgical procedures. Overall, the findings of this study contribute significantly to the field of NBTU, offering a robust framework for enhancing treatment precision and efficacy in clinical settings.

\end{abstract}

\begin{IEEEkeywords}

Finite Element Modeling, Pre-Operative Surgical Planning, Needle Based Therapeutic Ultrasound

\end{IEEEkeywords}

\section{Introduction}

\IEEEPARstart{I}ntracorporeal needle-based therapeutic ultrasound (NBTU) offers a minimally invasive option for the thermal ablation of malignant brain tumors, making it suitable for treating both primary and metastatic cancers \cite{gandomi3d}. Utilizing a high-frequency alternating electric field (up to 10 MHz) to excite a piezoelectric transducer, NBTU generates acoustic waves that propagate through tissue, causing localized high-temperature heating at the target tumor site. This rapid thermal elevation induces cell death, effectively ablating the tumor. While MR-guided laser interstitial thermal therapy (LITT) has shown experimental success in ablating radio-resistant brain metastases \cite{rao2014magnetic,carpentier2011laser}, NBTU provides a more precise ablation method by delivering lower doses of acoustic power directly to the tumor or targeted tissue, thereby sparing surrounding healthy tissue with greater efficiency \cite{ghoshal2013ex}. Moreover, NBTU demonstrates significant potential in treating otherwise inaccessible deep-seated tumors that are near or involve the vascular system \cite{missios2015renaissance,christian2014focused}.

To optimize the energy deposition and element design of NBTU transducers for effective thermal dose delivery during treatment, numerical modeling of the acoustic pressure field generated by the deforming piezoelectric transducer is a critical approach \cite{gandomi3d}. This modeling is often coupled with simulations of bioheat transfer processes to track the thermal propagation of the applicator over time. Magnetic resonance thermal imaging (MRTI) serves as an experimental validation method for these models, using the proton resonant frequency shift (PRFS) technique to derive quantitative spatial temperature maps from phase differences in the magnetic field. Validation studies using MRTI in gel phantoms have demonstrated the feasibility of these models and their ability to replicate thermal propagation patterns. However, for a more comprehensive evaluation of therapeutic efficacy, thermal damage isodose mapping is preferred. Building on our previous work \cite{gandomi2019thermo,gandomi2020modeling,gandomi3d,zhao2024deep}, this study aims to generate a spectrum of ablation data to support pre-operative surgical planning, providing a robust framework for predicting therapeutic outcomes and enhancing treatment precision.

\section{Tissue Damage Estimation}

Instead of using a measurement of the amount of energy delivered, thermal dose, which is a measurement of the amount of specific tissue damage caused by heat is widely used. Besides the CEM43 model used in \cite{szewczyk2022happens, zhao2024deep}, the Arrhenius model is also used in our model. In this model thermal damage is modeled as a first-order rate process in which tissue constituents transform from a native state to a damaged state with a reaction velocity (or rate constant), \emph{k}. The dependence of \emph{k} on temperature can be described by Arrhenius Equation \ref{equ:arrheniusch}: 

\begin{equation}
    k=Ae^{-\frac{E_a}{R(T+273)}}
    \label{equ:arrheniusch}
\end{equation}

Using this assumption the thermal dose, or thermal damage, $\Omega(\tau)$, is presented as Equation \ref{equ:damagech}

\begin{equation}
    \Omega(\tau)=\ln{\left(\frac{N(0)}{N(\tau)}\right)}=A\int_{0}^{\tau} e^{-\frac{E_a}{R(T+273)}} \,dt \label{equ:damagech}
\end{equation}

where $N(\tau)$ is the concentration of the native tissue at a heating duration of $\tau$, A is the pre-exponential factor (frequency factor), $E_a$ is the activation energy, \emph{R} is the universal gas constant, and $T$ is the absolute temperature, by adding 273 to make it in degrees of Kelvin. By using the Arrhenius model, tissue damage of both tumor and healthy brain tissue can be evaluated, so that more realistic modeling based on the tumor integrated brain ablation simulation can be achieved.

\section{Brain Tumor Integrated Simulation}

\subsection{Acoustic Pressure Map in the Trans-Tissue Medium}

We used a 3$\times$2mm (a$\times$b) oval shape acting as a tumor inside the brain. In the bioheat physics node additional bioheat material was added and selected tumor oval, then two thermal damage nodes were added under the brain and tumor bioheat material node, the properties parameters used in thermal damage and material properties are shown in Table \ref{tbl:tumor_medium}. The rest of the configuration was the same as described in \cite{zhao2024deep}.

\begin{table*}
    \centering
    \caption{Acoustic medium properties of healthy brain tissue, glioma, and meningioma. GM - gray matter. Data imported from \cite{sadeghi2016parameter,de2016heat,vaupel1989blood,bousselham2018brain, tanaka1965localization}.}
    \label{tbl:tumor_medium}
    \begin{tabular}{l|c|c|c|c}
    \hline
    Acoustic Medium Properties & Units &Brain (GM) & Glioma & Meningioma \\ \hline
    Heat Capacity ($C$)& $J/kg/^\circ C$ & 3680 & 3680 & 3680 \\ \hline
    Density ($\rho$)& $kg/m^3$ & 1035.5 & 1046 & 1058 \\ \hline
    Thermal Conductivity (\emph{K})& $W/m/^\circ C$ & 0.565 & 0.565 & 0.565 \\ \hline
    Blood Capacity ($C_{t,b}$)& $J/kg/^\circ C$ & 3630 & 3630 & 3630 \\ \hline
    Blood Density ($\rho_b$)& $kg/m^3$ & 1050 & 1050 & 1050 \\ \hline
    Blood Perfusion ($\omega_b$)& $kg/m^3/s$ & 0.013289 & 0.0005 & 0.0005 \\ \hline
    Speed of Sound (\emph{c})& $m/s$ & 1460 & 1660 & 1590 \\ \hline
    Metabolic Heat ($Q_m$)& $W/m^3$ & 16229 & 25000 & 25000 \\ \hline
    Frequency Factor ($A$) & $1/s$ & 1.66e91 & 3.153e47 & 1.737e91 \\ \hline
    Activation Energy ($E_a$)& $J/mole$ & 5.68e5 & 3.149e5 & 3.149e5 \\ \hline
    \end{tabular}
\end{table*}

\begin{figure}
    \centering
    \includegraphics[width=0.8\linewidth]{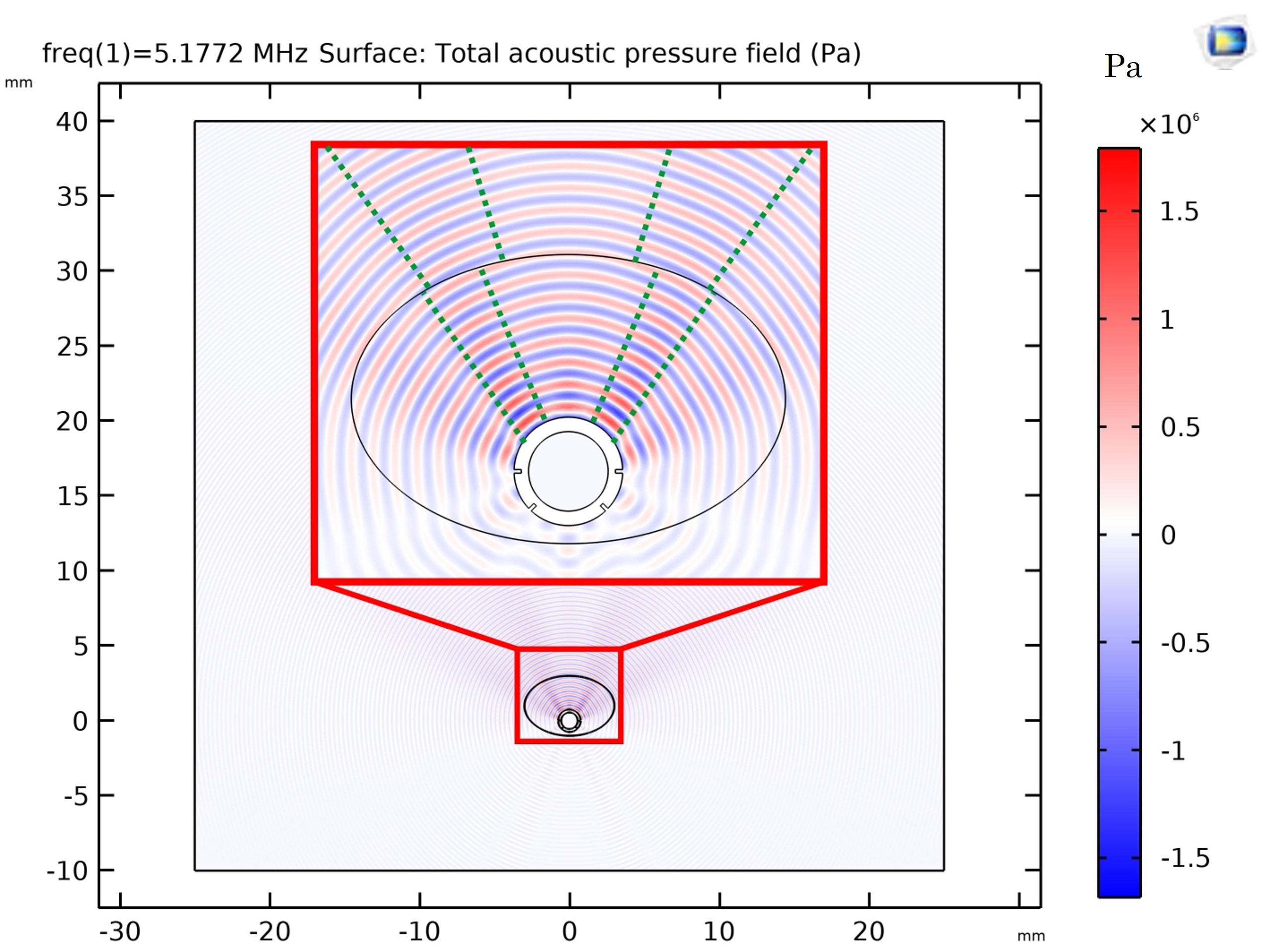}
    \caption{Acoustic pressure pattern of a 180$^\circ$ probe and glioma integrated medium.}
    \label{fig:acprtumor}
\end{figure}

We used a 180$^\circ$ probe and glioma integrated medium as an example, the acoustic pressure pattern can be found in Figure \ref{fig:acprtumor}, and the results show that the wave propagation was still following the pattern, however, there was a position shift and wave range width increase happened shown in green dashed lines symmetrically, and this was caused by the difference of brain and tumor proprieties. Although there was some misalignment with brain and tumor wave propagation, the pattern was still matching the patterns in our previous work \cite{campwala2021predicting,szewczyk2022happens,gandomi2019thermo,gandomi3d,gandomi2020modeling}. Based on the results, multiple configurations of medium and parameters simulation were achieved for more complex analysis. 

\subsection{Tumor Size Interference}

\begin{figure}
    \centering
    \includegraphics[width=0.95\linewidth]{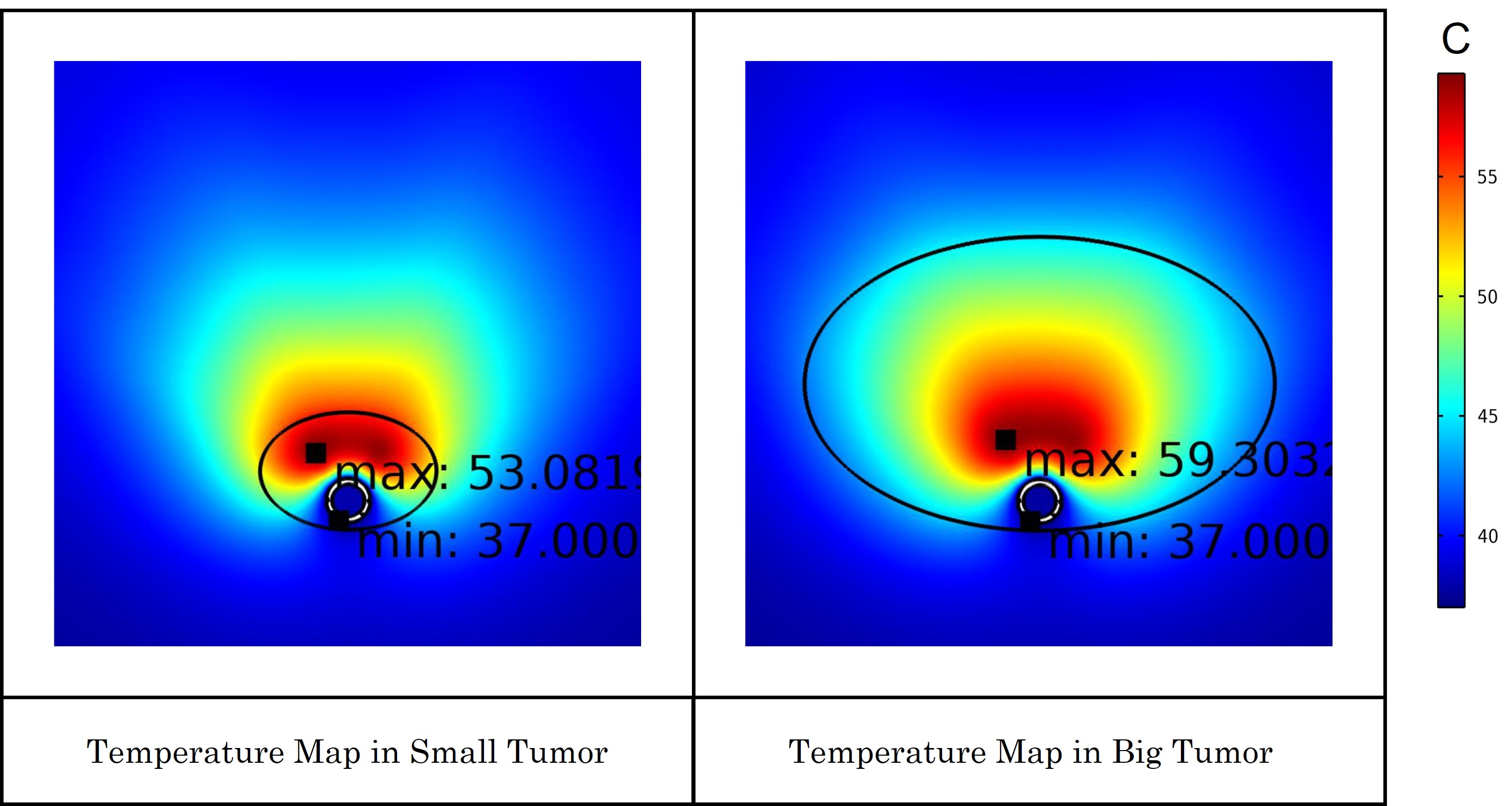}
    \caption{Thermal propagation maps in different sizes of tumors. The highest temperature reached is influenced by the size of the tumor.}
    \label{fig:Temp_BS}
\end{figure}

\begin{figure}
    \centering
    \includegraphics[width=0.95\linewidth]{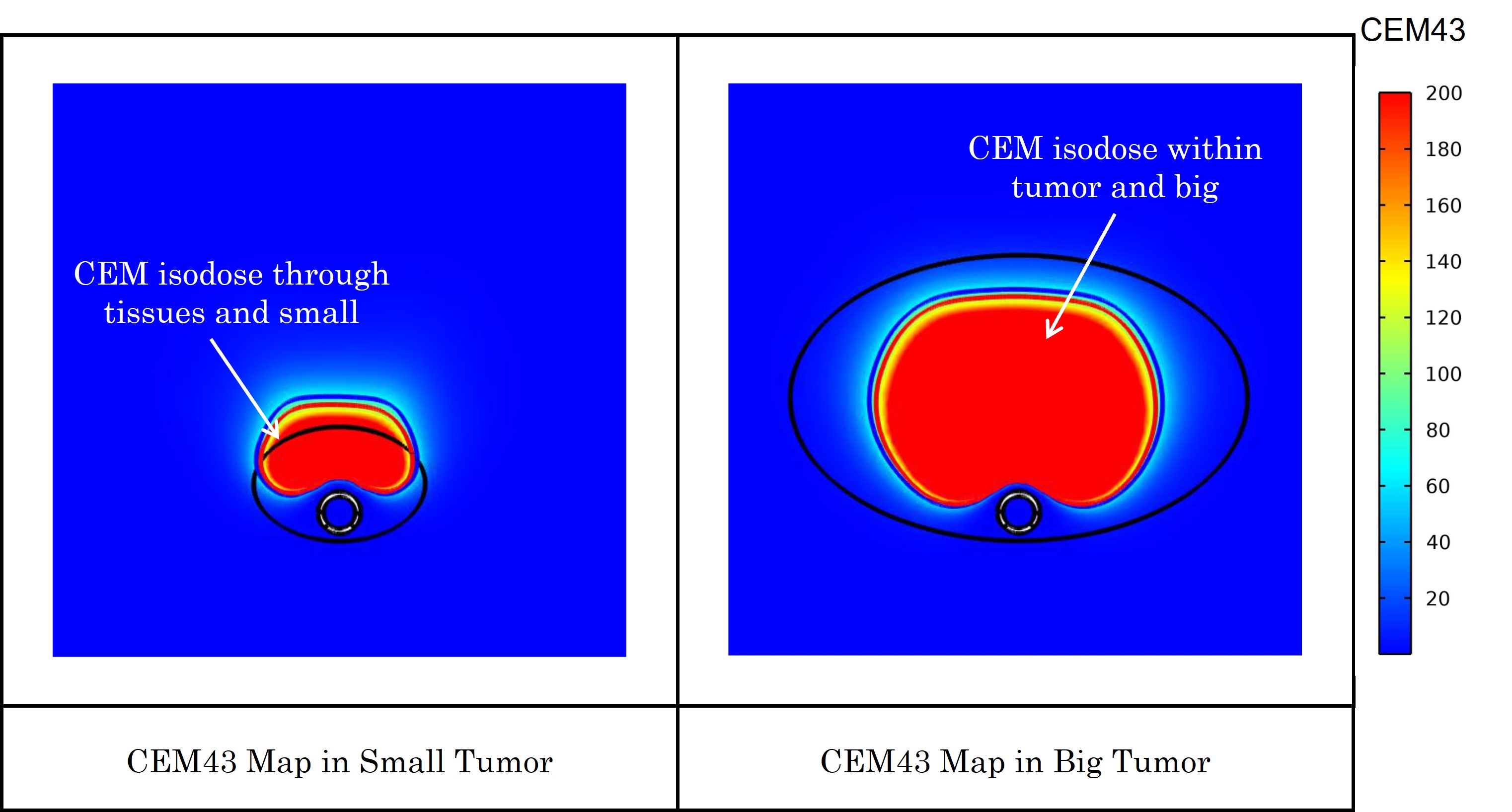}
    \caption{CEM43 isodose map in different sizes of tumors. The area of CEM43 70 and 100 is influenced by the size of the tumor.}
    \label{fig:cem_BS}
\end{figure}

The tumor size had a significant effect on the ablation process, which limited the thermal propagation through the tumor to the surrounding area. Figure \ref{fig:Temp_BS} shows the temperature propagation through the tumor with different sizes, namely 3$\times$2mm and 8$\times$5mm (a$\times$b) representing the small and large size of glioma, where the properties can be found in Table \ref{tbl:tumor_medium}, and with the same probe of the same ablation power, and same duration ablation condition. Results show the highest temperature and thermal range was reduced in the small tumor reaching up to 53.081$^\circ$C versus in the big tumor environment reaching up to 59.303$^\circ$C. By interfering with the temperature propagation, the thermal damage was influenced as well. Figure \ref{fig:cem_BS} shows the CEM43 model damage prediction in different sizes of tumors. The brain tissue and tumor CEM43 map were limited since the temperature was reduced. Because the wave propagation was not changed significantly, no shift or change was observed in the CEM43 of 70 and 100 isodose maps under the trans-tissue condition. However, the area size was changed, which is the same as the reduction of thermal range, in the small tumor the area of CEM43 isodose was reduced compared to the large tumor condition without a trans-tissue scenario. 

\begin{figure}
    \centering
    \includegraphics[width=0.95\linewidth]{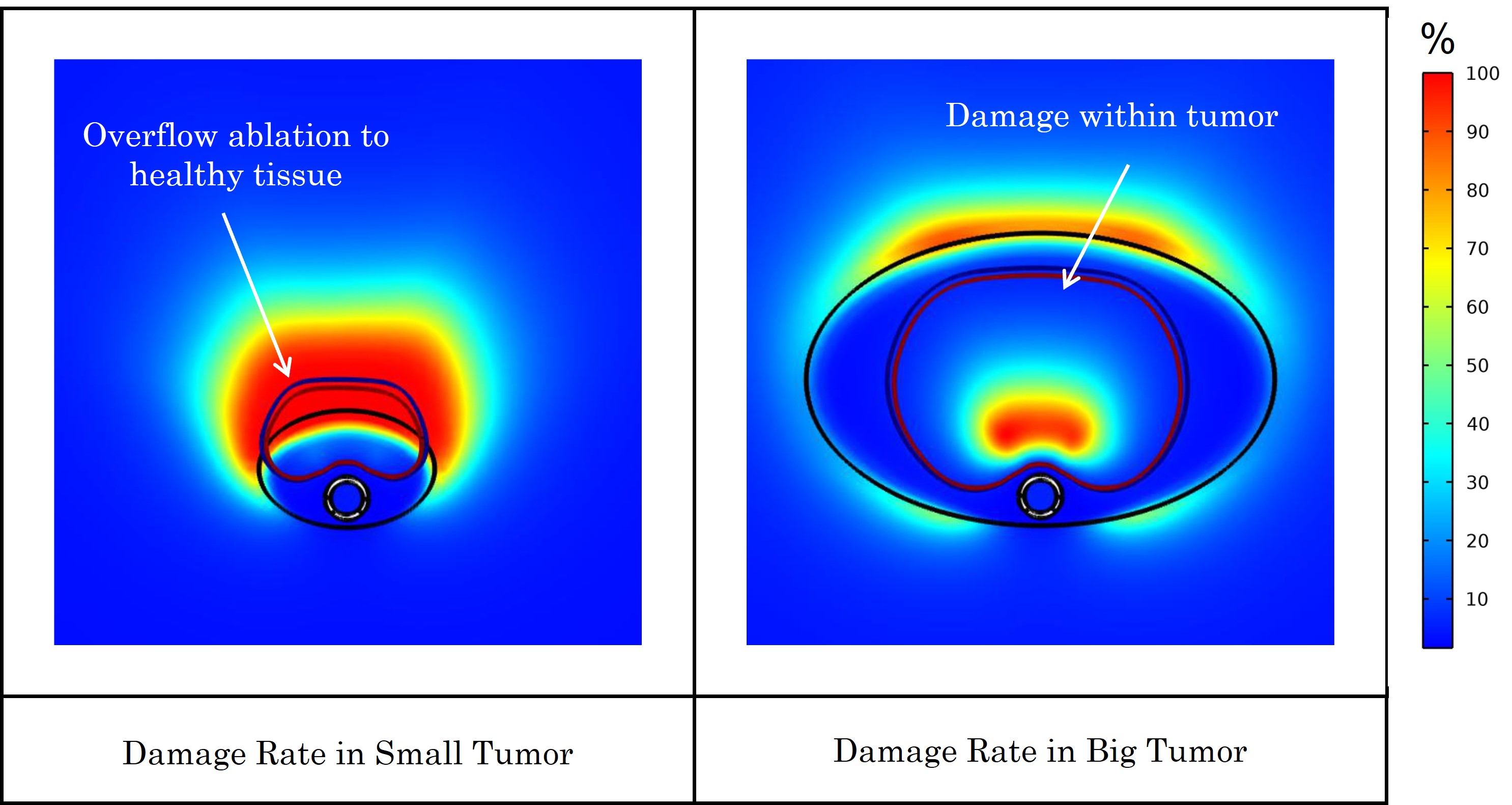}
    \caption{Tissue damage is estimated in different sizes of tumors. The small tumor is not ablated yet, the surrounding tissue already died, while for the big tumor, the damage was limited within the tumor range, however the surrounding tissue has started to heat up already, followed by unexpected ablation.}
    \label{fig:Damg_BS}
\end{figure}

The Arrhenius damage model was used in the model for further damage estimation, and the result of different size glioma ablation damage estimation can be found in Figure \ref{fig:Damg_BS}. Results show a very bad condition of over-ablated to healthy tissue with the same probe of the same ablation power and same duration ablation condition. Especially in small tumor conditions, the tumor was not ablated yet, while the surrounding healthy tissue was already unexpectedly ablated. In a big tumor, the damage was limited within the tumor range, however the surrounding tissue has started to heat up already, followed by unexpected ablation.

\subsection{Tumor Type Interference}

\begin{figure}
    \centering
    \includegraphics[width=0.95\linewidth]{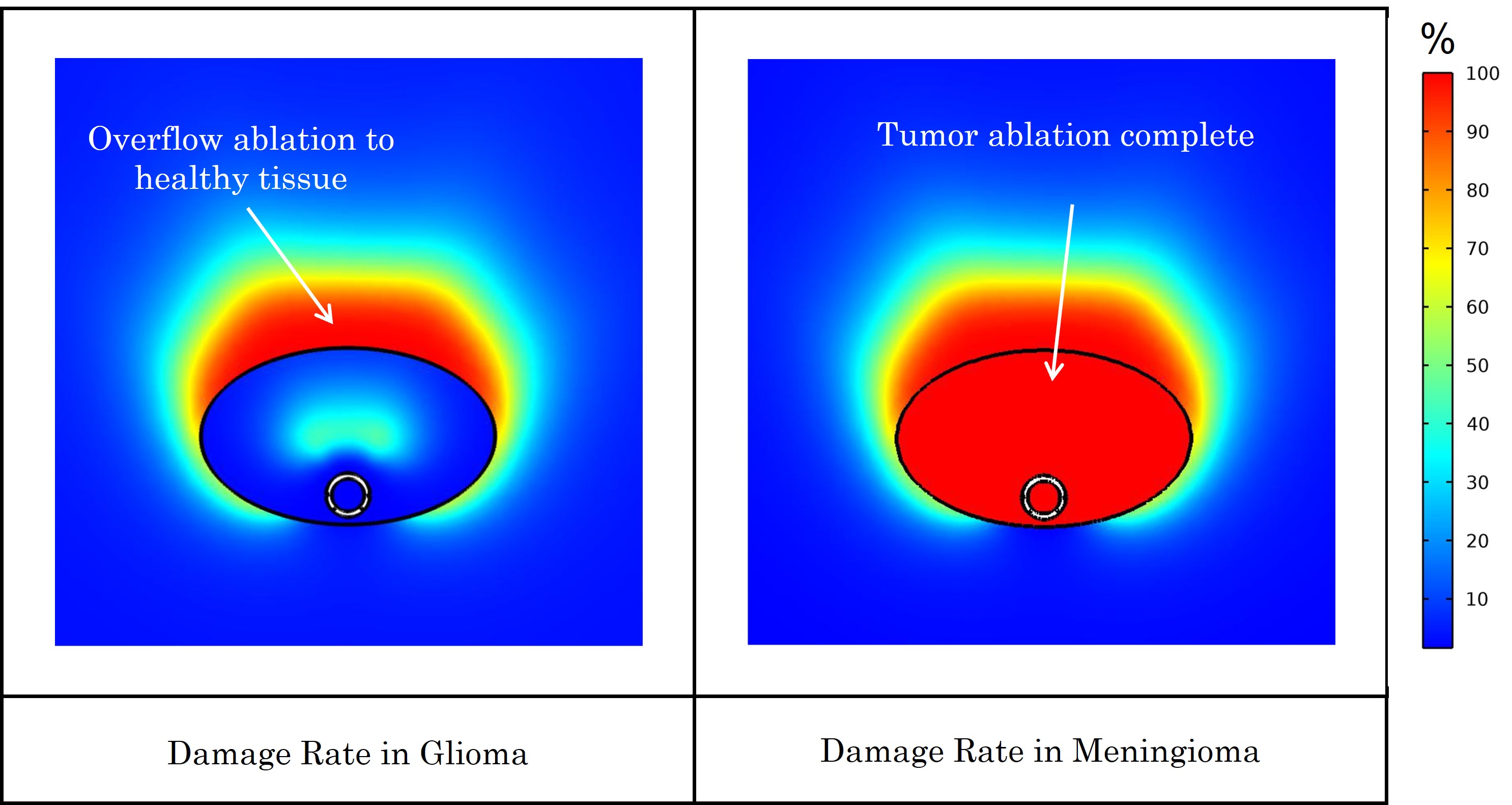}
    \caption{Tissue damage is estimated in different types of tumors, namely glioma and meningioma.}
    \label{fig:damage_tumor_type}
\end{figure}

Besides the size of tumor interference, the type of tumor also influences the ablation performance of the different properties between different tumor types. In this study, we compared two common brain tumors, namely glioma and meningioma considering these two types of tumor varies significantly in pre-exponential factor ($A$). Results of tissue damage are shown in Figure \ref{fig:damage_tumor_type}, which are also with the same probe of the same ablation power and same duration ablation condition. It is observed that because the meningioma has a larger A factor compared to the glioma, the damage of ablation is more complete, while the glioma is not active enough so the over ablated condition happened in this condition. To avoid a mixed situation, the trans-tissue should be reduced, which indicates that the pre-surgical estimation and planning of ablation should be accurate enough to ablate the target tumor homogeneous medium edge only.

\section{Thermal Ablation Probe Parameter Simulation}
\label{sec:c7s4}

We have introduced the tumor integrated ablation simulation, and some major conditions that affected the ablation performance were discussed such as the tumor size and tumor type, which indicated that it is necessary to select the proper probe parameters and there is still room for optimization. In this section, we used the simulation model to achieve the spectrum of ablation performance with parameters combination which provides the database preparation for further studies.

\begin{figure}
    \centering
    \includegraphics[width=0.8\linewidth]{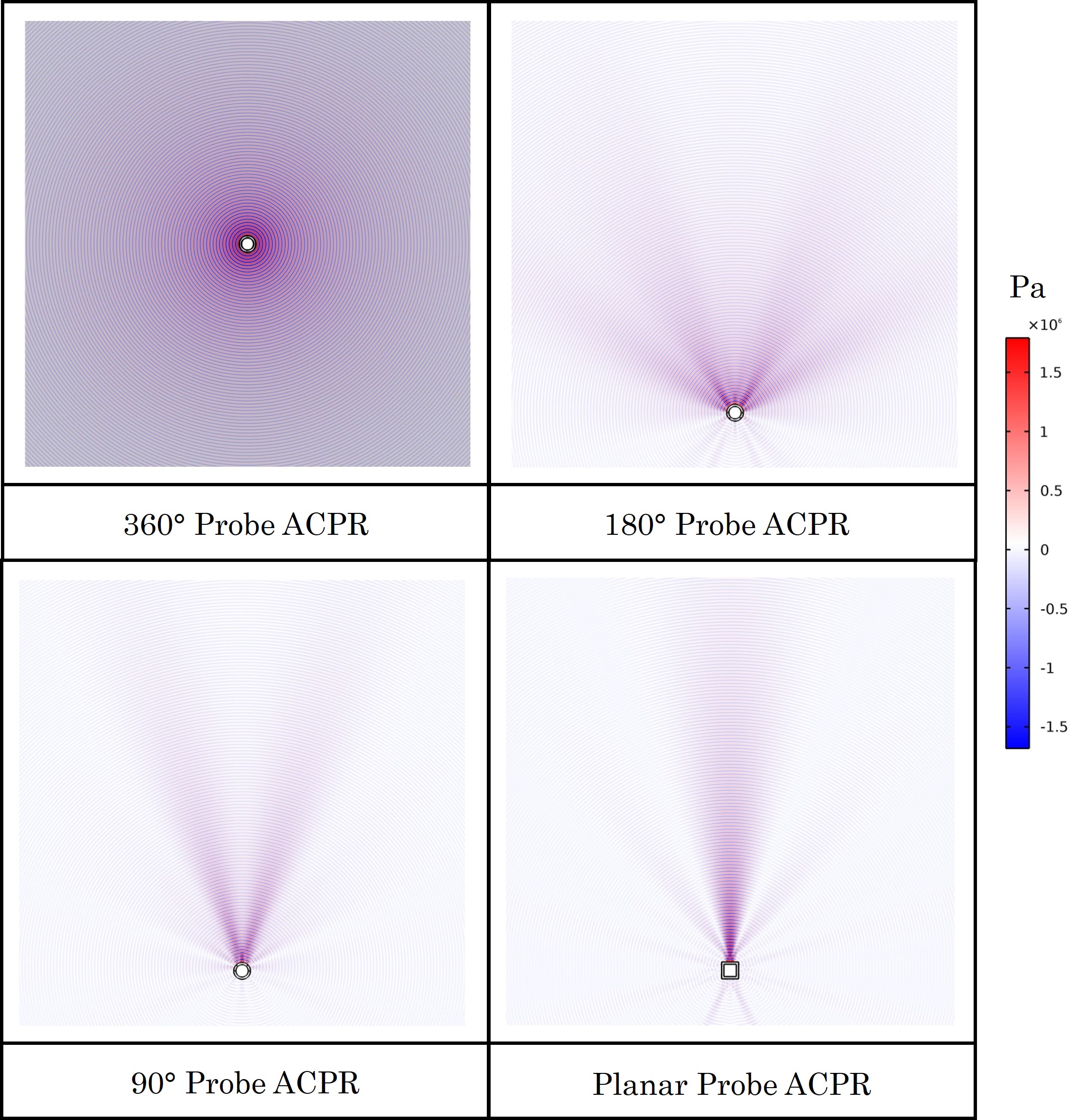}
    \caption{Acoustic pressure pattern from the 4 most common directional probes used in the study. }
    \label{fig:4acpr}
\end{figure}

\begin{figure*}
    \centering
    \includegraphics[width=0.9\linewidth]{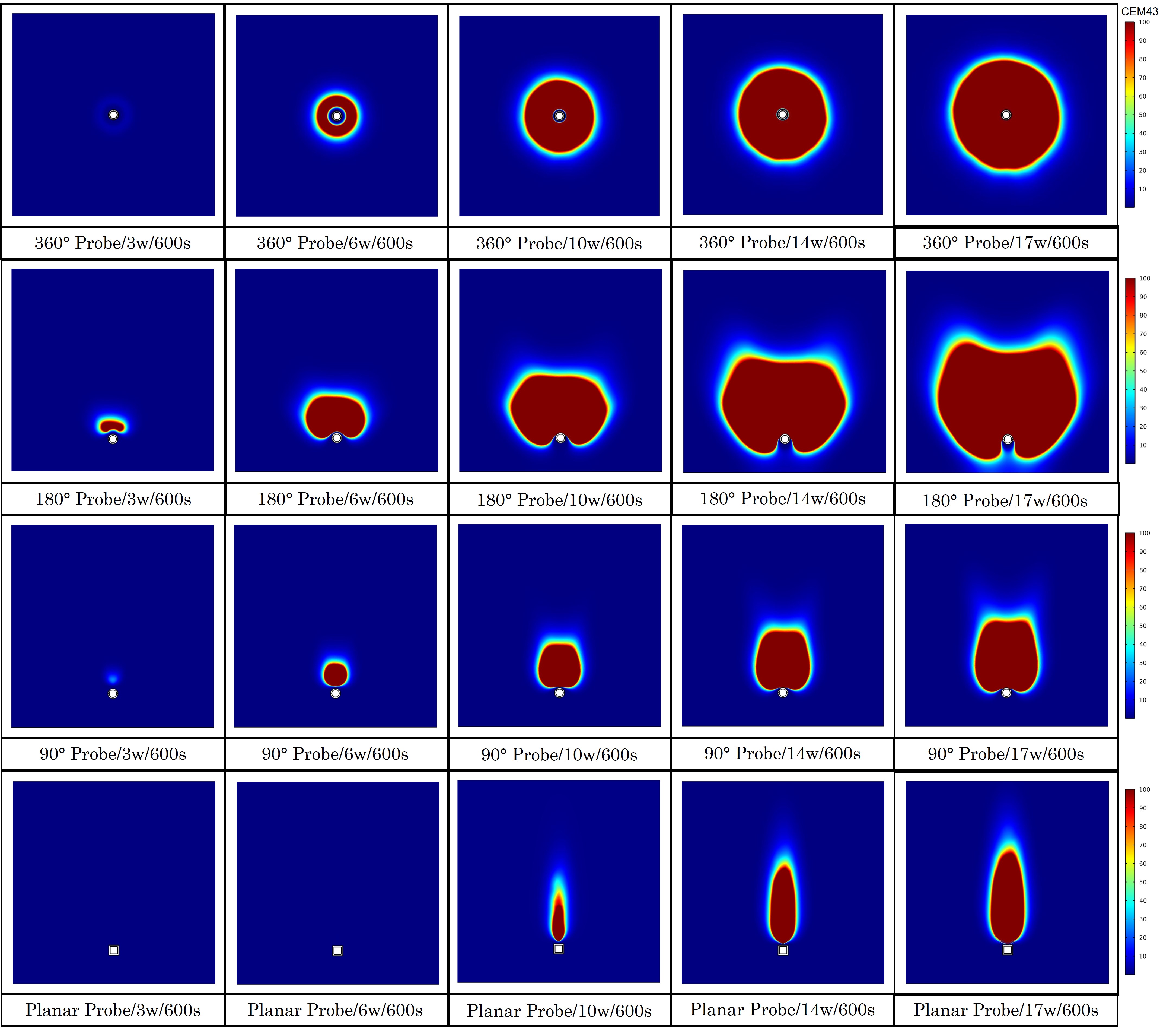}
    \caption{CEM43 isodose maps of 360$^\circ$, 180$^\circ$, 90$^\circ$, and planar probe under 3W, 6W, 10W, 14W, and 17W acoustic power respectively.}
    \label{fig:CEM_pattern}
\end{figure*}

\begin{figure*}
    \centering
    \includegraphics[width=0.75\linewidth]{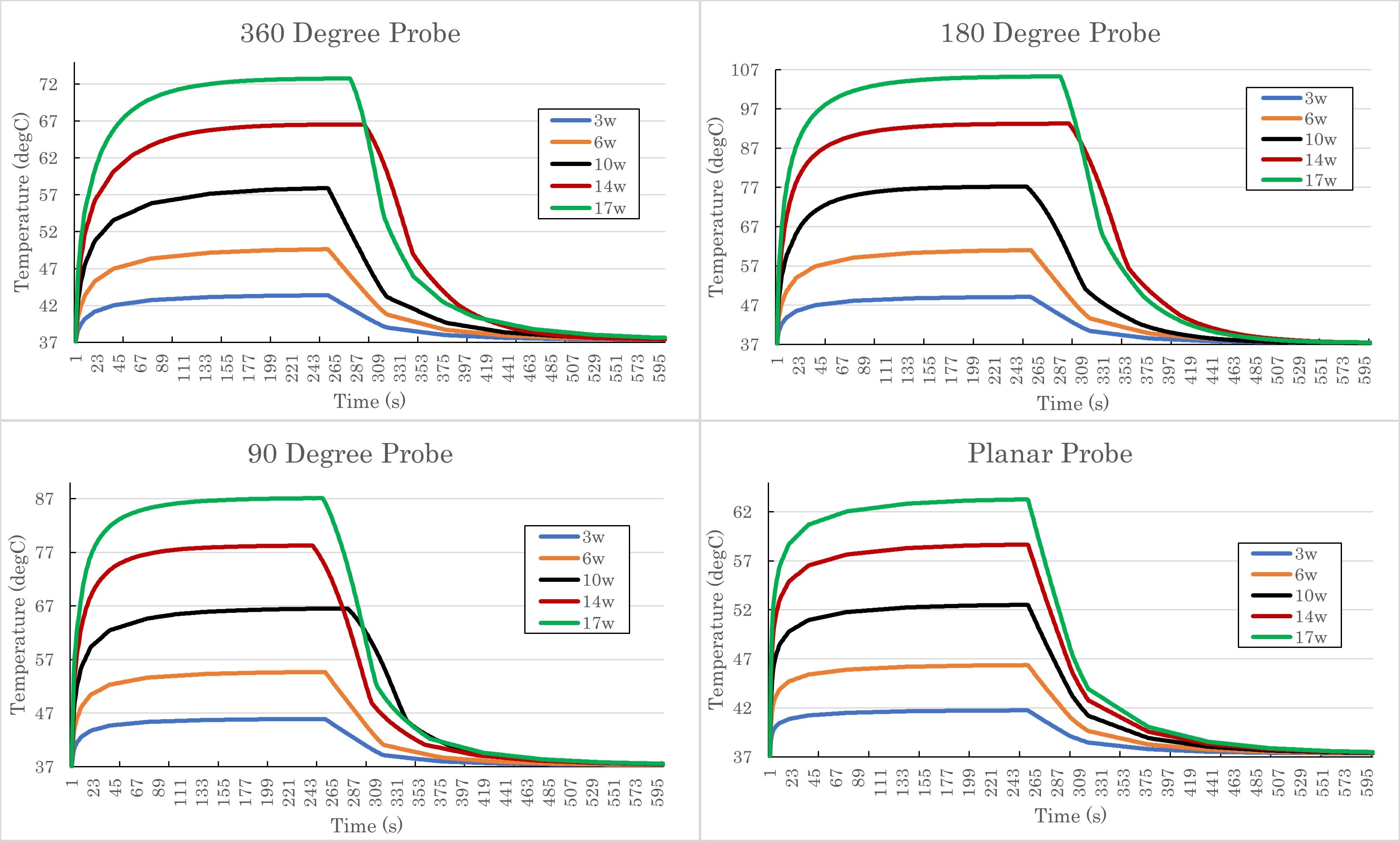}
    \caption{The highest temperature versus time for all the simulation conditions with different types of probes and acoustic power.}
    \label{fig:tempall}
\end{figure*}

In this study, we chose the four most commonly used directional probes, namely 360$^\circ$, 180$^\circ$, 90$^\circ$, and planar probe with a square tube transducer, and for each probe, we chose 3W, 6W, 10W, 14W and 17W acoustic power under 600s (300s probe on and 300s cooling down) duration time configuration to estimate the trends for our ablation spectrum construction. Following the same analysis procedure, we first achieved the acoustic pressure pattern map, which is shown in Figure \ref{fig:4acpr}. Note that all the parameter used for the acoustic medium was based on glioma, and the power was 10W for all the figures shown above. Based on the different power of acoustic pressure pattern maps, we summarized 20 CEM43 model thermal damage isodose maps in glioma, and the results are shown in Figure \ref{fig:CEM_pattern}. For different types of probes, the highest temperature reached was proportional to the power used and also varied with probe type. Figure \ref{fig:tempall} shows the time used to reach the highest temperature for all the simulation conditions with different types of probes and acoustic power. We can observe that the highest temperature reached was related to the probe type under 17W largest acoustic power setup, where the 180$^\circ$ probe reached the highest value of 105.29$^\circ$C, followed by 90$^\circ$ probe with 87.08$^\circ$C, 360$^\circ$ probe with 72.78$^\circ$C, and planar probe with 62.83$^\circ$C, and reduced power will lower the temperature reached. Moreover, the time used to reach the max temperature was different, where the planar probe used the shortest time of approximately 80s to reach the low increasing period, followed by 90$^\circ$ probe with 90s, 180$^\circ$ probe with 100s, and 360$^\circ$ probe with 150s. All these data are solid preparation for later probe selection protocol construction.

\section{Probe Parameters Selection Method for Surgical Pre-Planning}

Instead of using the temperature propagation method, the thermal damage model is more important for pre-surgical planning work. In this section, we developed a probe parameters selection method for surgical pre-planning with surgeons and researchers with clinical usage. 

Based on the results shown in Figure \ref{fig:CEM_pattern}, the pattern of different probes varied a lot with each other, and the selection of directional probe parameters of probe type, power used, and trajectory planning played a vital role in the pre-surgical planning procedure. To address the precise ablation procedure, the following simulation data was accumulated.   

\begin{figure*}
    \centering
    \includegraphics[width=0.7\linewidth]{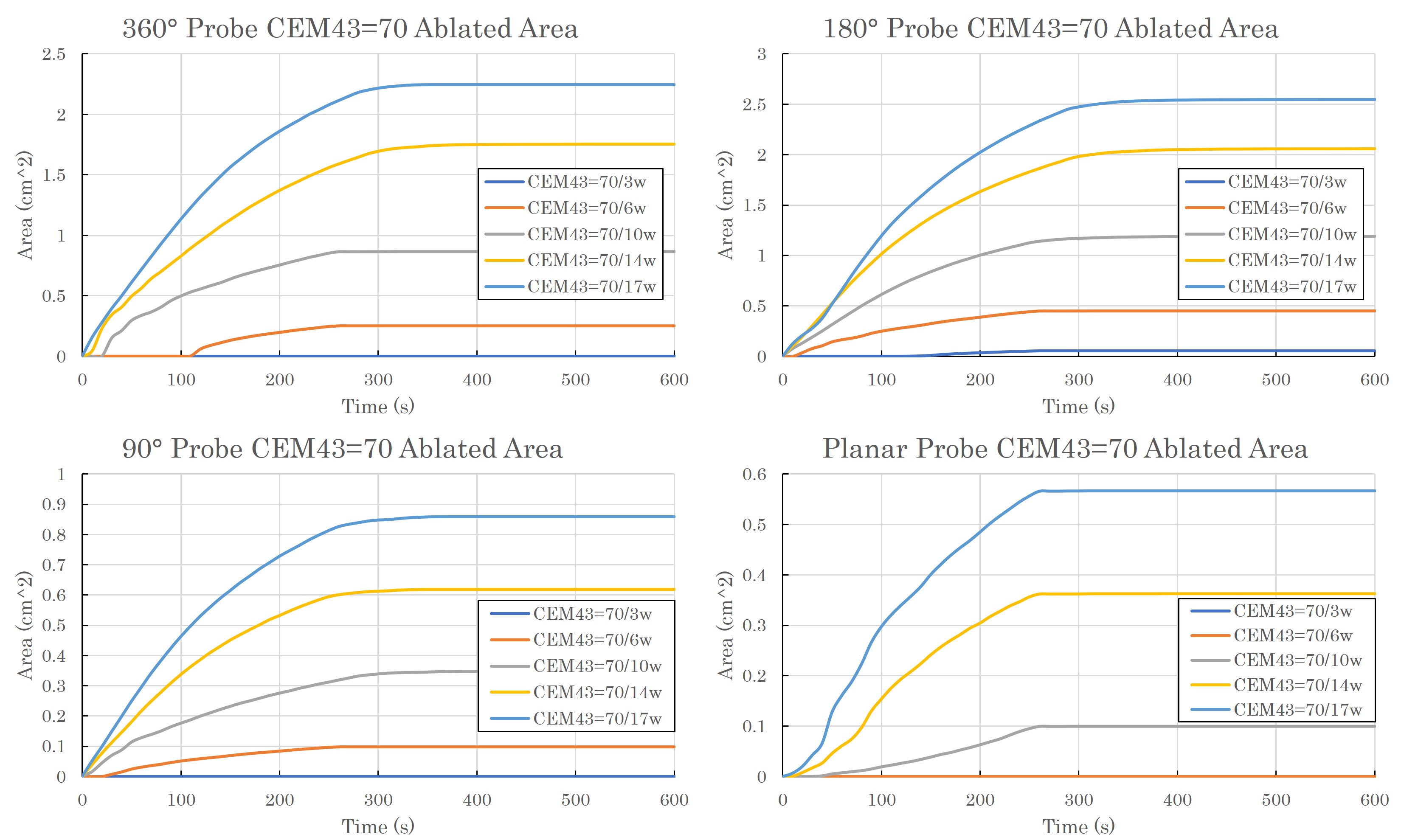}
    \caption{The CEM43 of 70 isodose areas versus time for all the simulation conditions with different types of probes and acoustic power.}
    \label{fig:cem70_ablate_time}
\end{figure*}

\begin{figure*}
    \centering
    \includegraphics[width=0.7\linewidth]{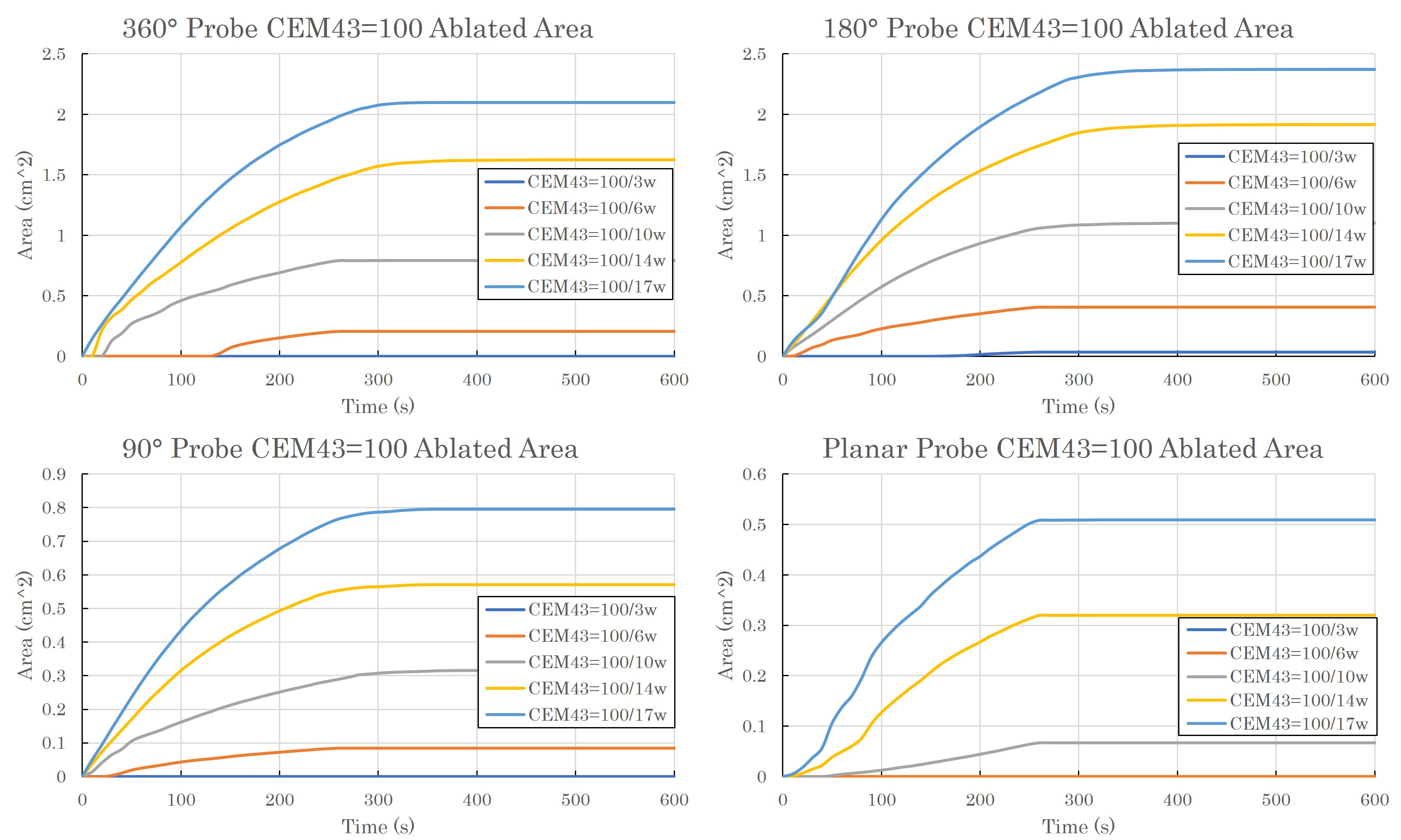}
    \caption{The CEM43 of 100 isodose area versus time for all the simulation conditions with different types of probes and acoustic power.}
    \label{fig:cem100_ablate_time}
\end{figure*}

\begin{figure*}
    \centering
    \includegraphics[width=0.7\linewidth]{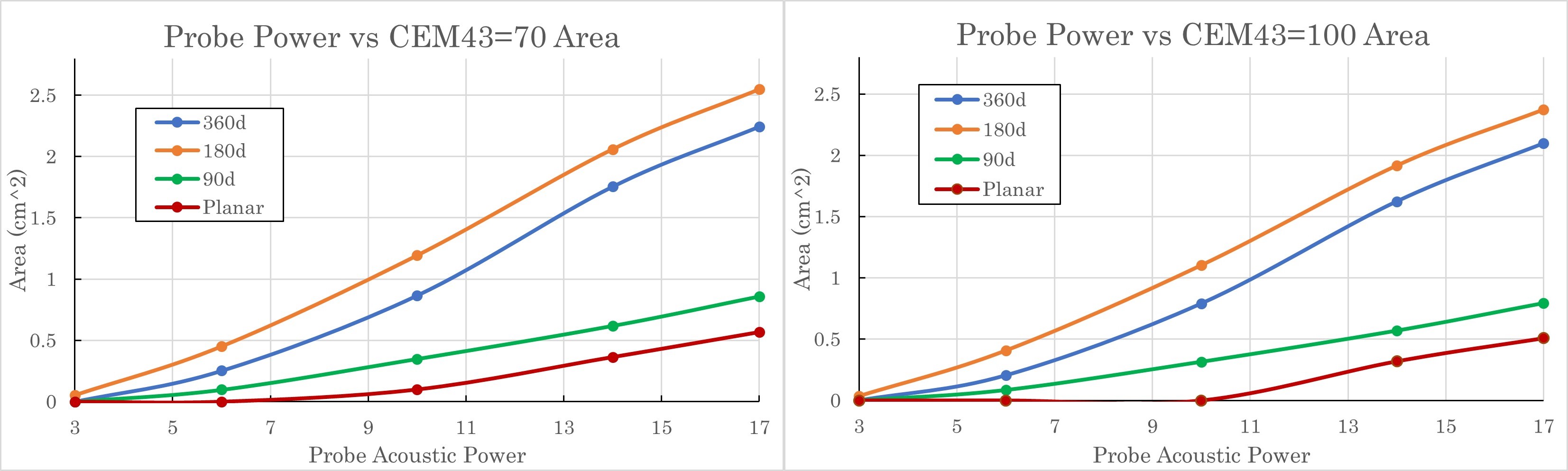}
    \caption{Acoustic power versus CEM43 of 70 and 100 largest ablated isodose area under 600s duration time with four types of probe.}
    \label{fig:cem_area}
\end{figure*}

\begin{figure*}
    \centering
    \includegraphics[width=0.7\linewidth]{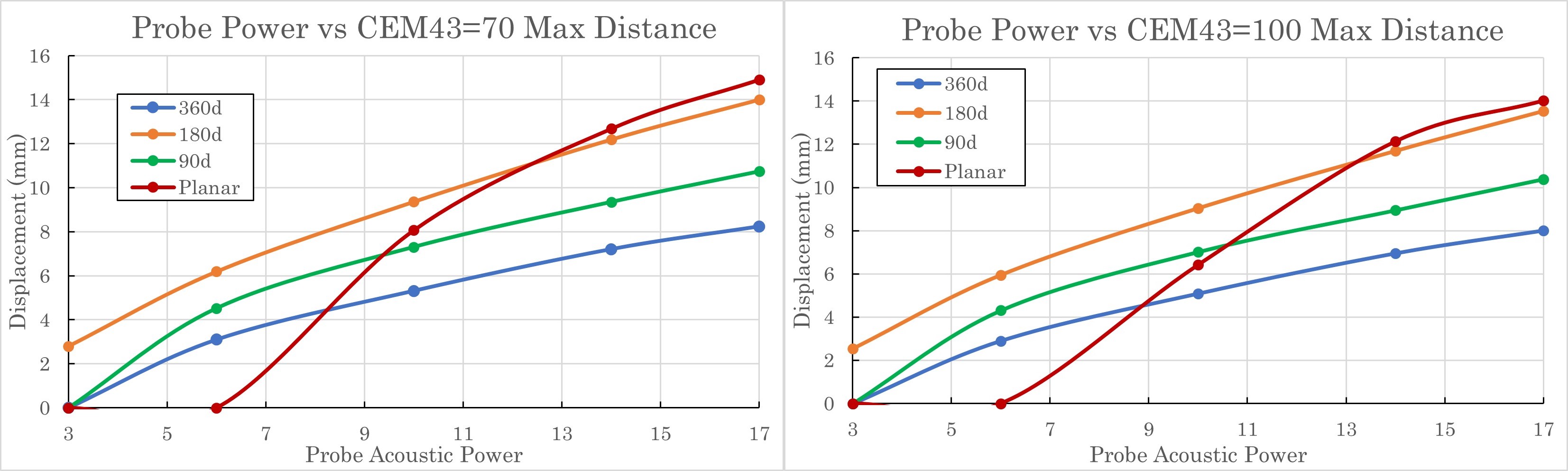}
    \caption{Acoustic power versus CEM43 of 70 and 100 maximum ablated isodose distance under 600s duration time with four types of probe.}
    \label{fig:cem_dis}
\end{figure*}

Figure \ref{fig:cem70_ablate_time} and Figure \ref{fig:cem100_ablate_time} show the CEM43 of 70 and 100 isodose area versus time for all the simulation conditions with different types of probes and acoustic power, which indicates the area of ablation versus time data. This is an area that gradually increases the trend for different probe parameter combinations. Moreover, Figure \ref{fig:cem_area} and Figure \ref{fig:cem_dis} show the acoustic power versus CEM43 of 70 and 100  ablated isodose largest area and maximum distance respectively under 600s duration time with four types of probe. This is the most important and valuable database in this work, which acts as a spectrum dictionary and can be used both in the probe selection protocol and further smart ablation studies.

\begin{figure}
    \centering
    \includegraphics[width=0.8\linewidth]{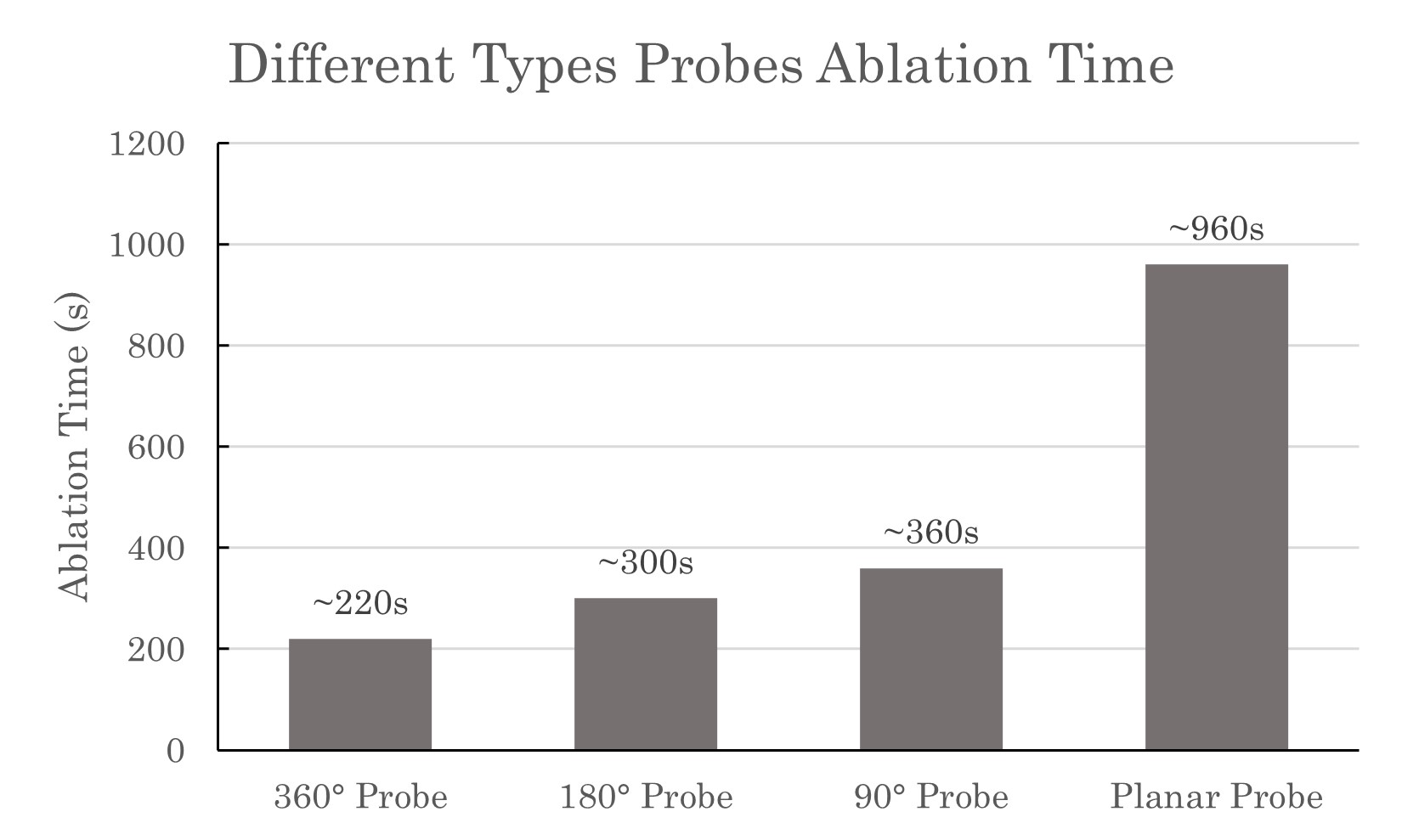}
    \caption{Ablation complete-time was used from four types of directional probes, under a 5mm radian round sample tumor pattern.}
    \label{fig:ablation_time}
\end{figure}

\begin{figure}
    \centering
    \includegraphics[width=0.8\linewidth]{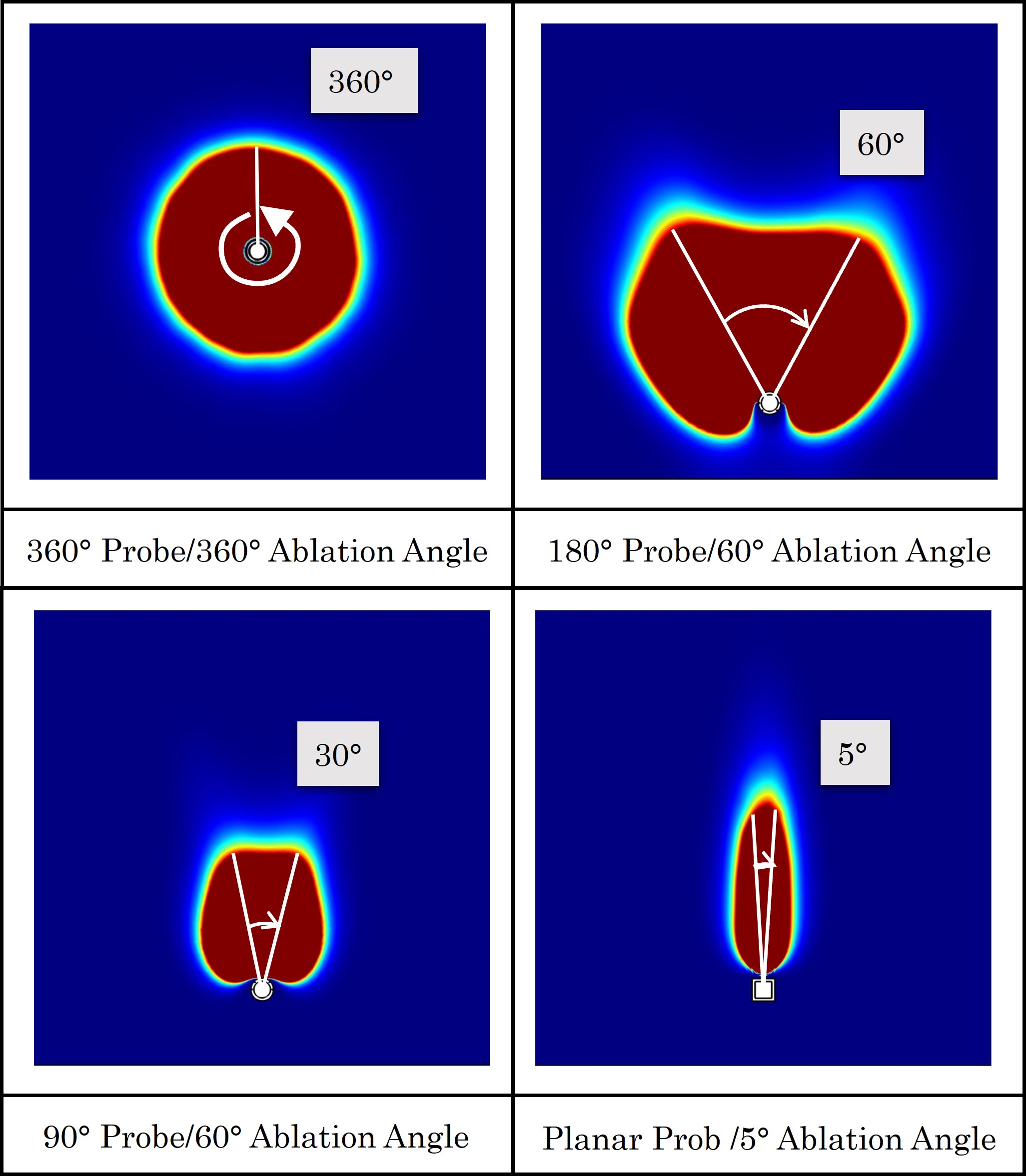}
    \caption{Ablation angle ($\theta$) generated by four types of directional probes, namely 360$^\circ$ for 360$^\circ$ probe, 60$^\circ$ for 180$^\circ$ probe, 30$^\circ$ for 90$^\circ$ probe, and 5$^\circ$ for the planar probe.}
    \label{fig:ablation_angle}
\end{figure}

We built a dummy tumor with a 5mm radius as a sample pattern to calculate the approximate ablation time based on all the simulation results we had accumulated. It is observed from Figure \ref{fig:ablation_time} that the 360$^\circ$ probe performed the shortest time at approximately 220s compared to other types of probes, namely followed by 180$^\circ$ probe with 300s, 90$^\circ$ probe with 360s, and planar probe with 960s. Figure \ref{fig:ablation_angle} shows the Ablation angle ($\theta$) generated by four types of directional probes, namely 360$^\circ$ for 360$^\circ$ probe, 60$^\circ$ for 180$^\circ$ probe, 30$^\circ$ for 90$^\circ$ probe, and 5$^\circ$ for the planar probe. Combining the two results above, we can get a summarized characteristic of the four types of probes in this study. The 360$^\circ$ generates a round pattern with the largest pattern range and ablation angle, it is suitable for a round or oval shape tumor and performs the fastest ablation time, however, it is not suitable for the irregular shape of tumors with sharp spikes. On the other hand, the 90$^\circ$ probe generates a flame shape of ablation pattern, which provides the smallest and finest tips of ablation capability, or the highest resolution of ablation, however, it uses the longest time of ablation which is opposite to the 360$^\circ$ probe. The 180$^\circ$ and 90$^\circ$ probe are between the characteristics. A comparison summary of key parameters from four types of probes can be found in Table \ref{tbl:paracomp}. In this study, the time used for ablation is the priority to be considered because a fast procedure improves the patient outcome significantly, so regardless of the power used during the ablation procedure, we suggest choosing a larger degree probe as much as possible as it can be used.

\begin{table}[]
    \centering
    \caption{Key parameters of four types of probes compare summary.}
    \label{tbl:paracomp}
    \begin{tabular}{c|c|c|c}
    \hline
    Probe & Range & Time & Resolution\\ \hline
    360$^\circ$ & Large & Fast & Low  \\ \hline
    180$^\circ$ & Moderate & Mid-Fast & Moderate  \\ \hline
    90$^\circ$  & Mid-Small & Moderate & Mid-high  \\ \hline
    Planar      & Small & Slow & High \\ \hline
    \end{tabular}
\end{table}

\begin{figure}
    \centering
    \includegraphics[width=0.8\linewidth]{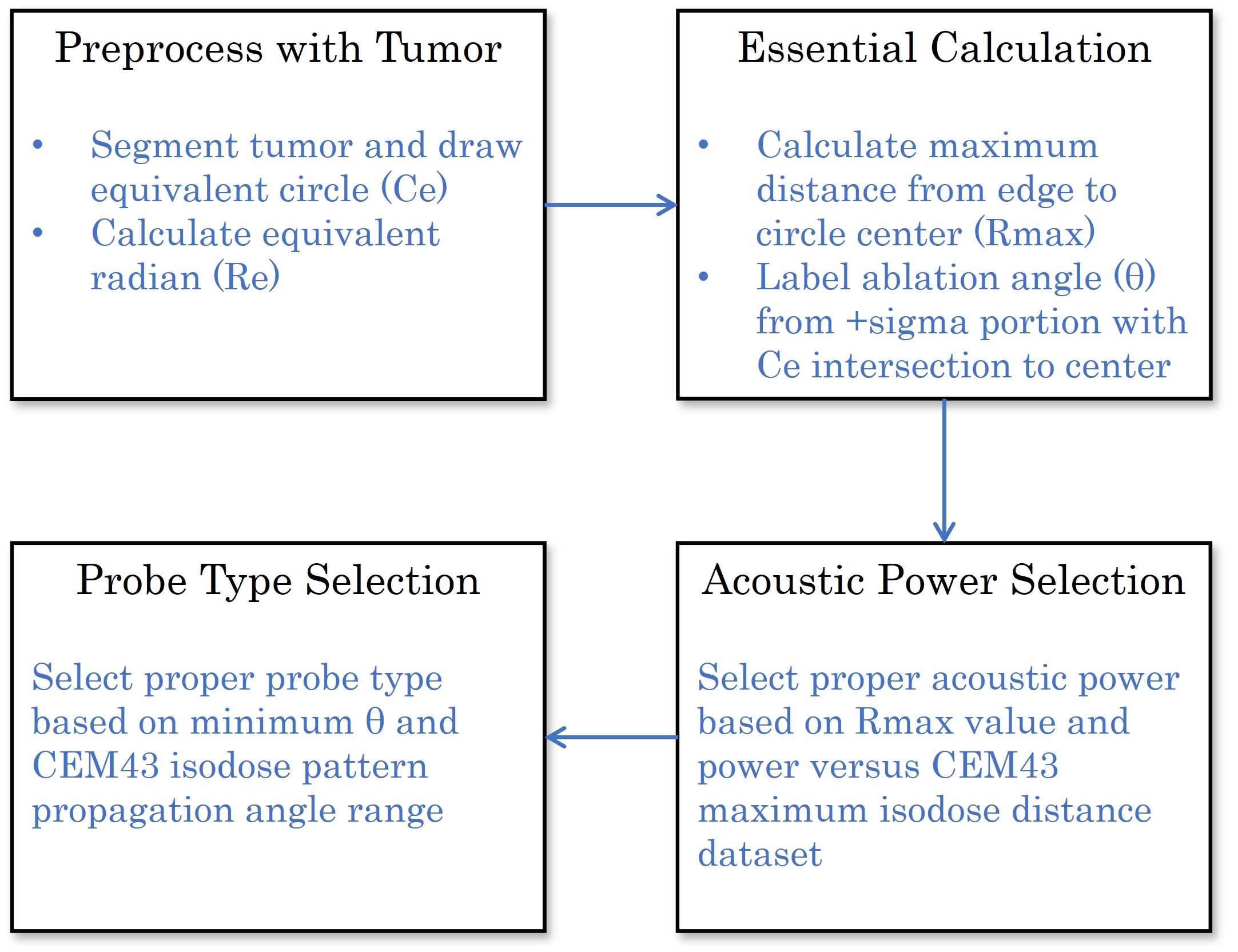}
    \caption{Probe selection protocol for pre-surgical planning.}
    \label{fig:protocol}
\end{figure}

\begin{figure}
    \centering
    \includegraphics[width=0.75\linewidth]{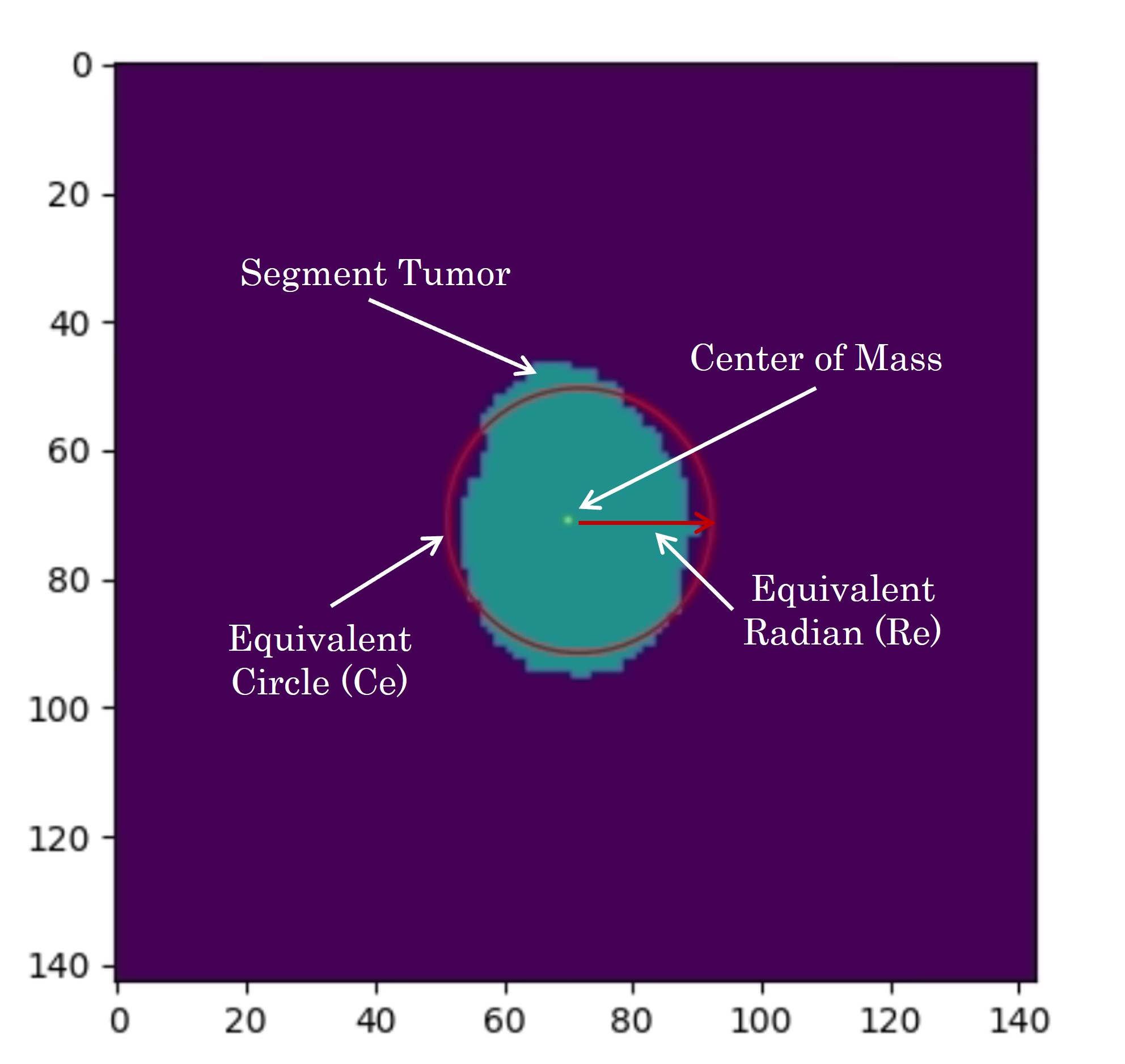}
    \caption{Segment the tumor scan images, draw the equivalent circle, and label the circle center.}
    \label{fig:segment}
\end{figure}

Based on the key parameters compared summary and all the simulation data we have accumulated, a preliminary probe selection protocol for pre-surgical planning towards ablation procedure was developed, which is shown in Figure \ref{fig:protocol}. We first segment the tumor from the scanning images and draw an equivalent circle ($C_e$), this circle is generated by counting the pixels of the tumor portion to get the area value, then using the center of mass as the circle center equivalent $O$, this equivalent $O$ is the probe insertion target point. The preprocess with tumor is shown in Figure \ref{fig:segment}. The equivalent radian $R_e$ can also be calculated, as well as the internal and external portion compared to the $C_e$, namely $-\sigma$ and $+\sigma$. Note that the internal portion of $-\sigma$ can be achieved by decreasing the power during the procedure to reduce the ablation distance. Then some essential parameters are calculated and labeled, namely the maximum distance ($R_{max}$) from the edge with the largest $+\sigma$ value to equivalent center $O$ following Equation \ref{equ:rmax}, and the ablation angle ($\theta$). Note that for compatibility consideration, we choose the minimum ablation angle for this step, namely $\theta_{min}$. The next step is the acoustic power selection, which is based on the $R_{max}$ value and looks for compatible acoustic power from all four types of probes from the power versus CEM43 maximum isodose distance data set (Figure \ref{fig:cem_dis}). The last step is to finalize the probe, which is based on the $\theta_{min}$ and the CEM43 isodose pattern propagation angle range (Figure \ref{fig:ablation_angle}). In this step, we will choose the probes with an ablation range angle that is smaller than $\theta_{min}$ so that no over ablation issue will happen. Note that if multiple probes are still under consideration after all the protocols, choose the largest ablation angle range probe for the final decision.

\begin{equation}
    R_{max} = R_e + (+\sigma)
    \label{equ:rmax}
\end{equation}

\begin{figure}
    \centering
    \includegraphics[width=0.95\linewidth]{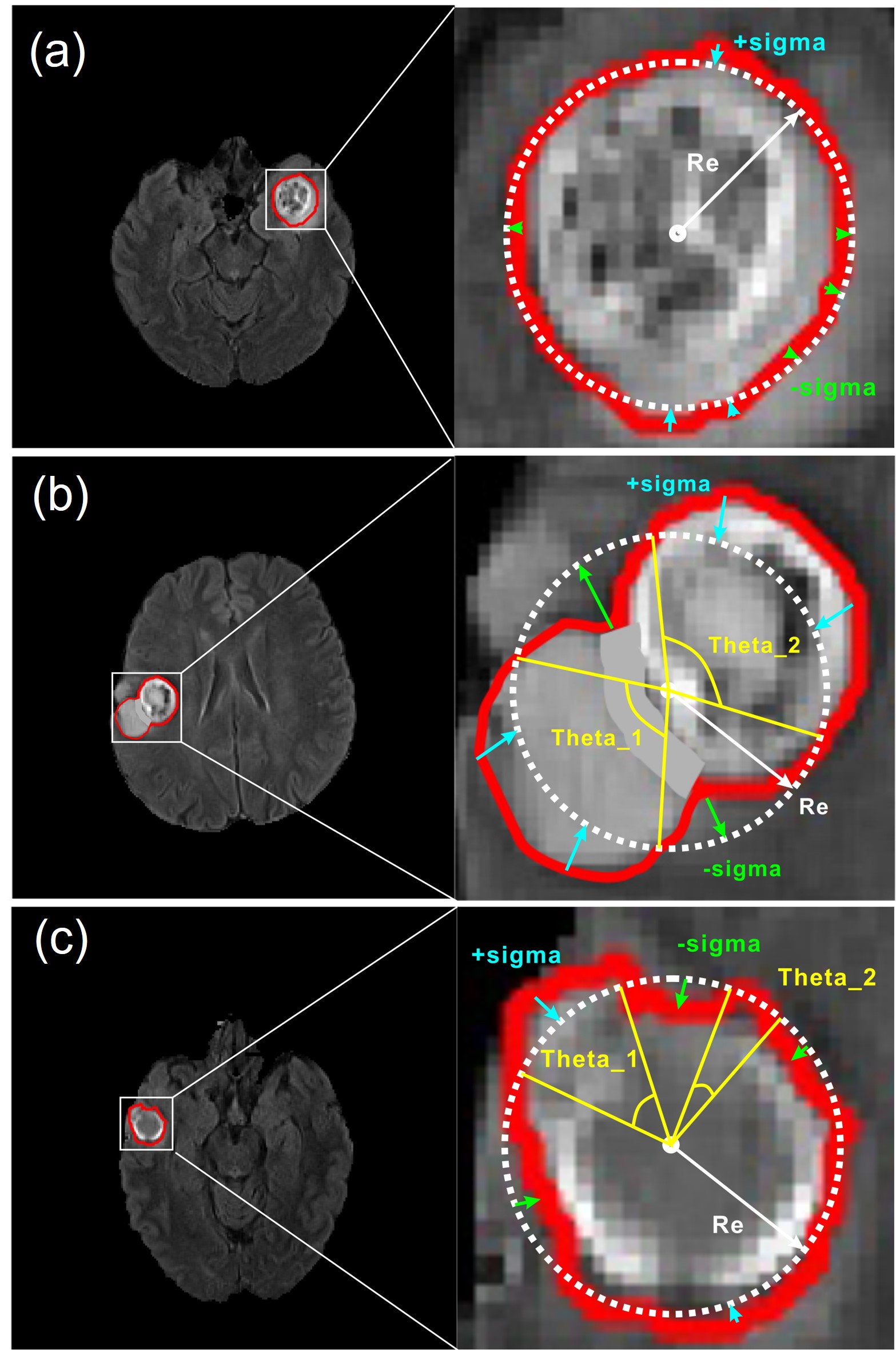}
    \caption{3 tumor samples after labeled with $R_{max}$, $\pm\sigma$ portions, and $\theta_{min}$}
    \label{fig:tumor_sample}
\end{figure}

Figure \ref{fig:tumor_sample} shows 3 samples of tumor after labeled with $R_e$, $\pm\sigma$ portions, and $\theta_{min}$. (a) the tumor is almost a circle with very small $\pm\sigma$ portions, so this tumor is suitable with a 360$^\circ$ probe for rapid ablation. (b) the tumor is a $H_2$ molecule shape, although labeled with essential parameters, and the $\theta_{min}$ is larger than 60$^\circ$, which means it is suitable with 180$^\circ$ probe, however, this pattern is more suitable for multiple entry points insertion with multiple ablation duration. This is not in our study and requires further optimization study. And lastly (c) tumor is an irregular shape with two $\theta$ values namely approximately 45$^\circ$ and 10$^\circ$. Considering the 10$^\circ$ of $+\sigma$ portion is very small and the edge is almost located on the edge without a large spike compared to the 45$^\circ$ portion, we can choose the 90$^\circ$ probe for rapid ablation or the planar probe for the high resolution ablation procedure. 

\section{Conclusion and Disscussion}

This work presents the database preparation for future studies based on the extended simulation results. We first developed an extended simulation integrated with different tumors to evaluate the effect of thermal damage in trans-tissue conditions. Then we accumulated the simulation database of four directional probes, including the CEM43 isodose map, temperature change, thermal dose area, and maximum distance. Finally, an idea of the probe selection method was presented for pre-surgical planning.

In the tumor integrated simulation, we involved the Arrhenius equation for tissue damage estimation, and results showed a significant effect from the tumor size and type with the trans-tissue scenario, so we suggested avoiding thermal propagation and damage model with trans-tissue modeling, as well as in the real clinical surgery. Also, the change of parameters during the ablation process should also be considered, especially when dealing with multiple tissues in a complex environment.

Then the database of four directional probes, including the CEM43 isodose map, temperature change, thermal dose area, and maximum distance based on the simulation model built up the dictionary of thermal ablation therapy, this work will benefit the future complex trajectory planning. Furthermore, this simulation method can also build up an easy-to-use platform for researchers to develop various parameters in combination with NBTU thermal ablation study.

Finally, the idea of the directional probe selection method gave us inspiration for how to use the database developed above, and this method will also benefit the pre-surgical planning study and eventually reduce the study curve for surgeons and researchers. The idea of choosing a large angle probe will significantly improve the outcomes of patients by reducing the ablation procedure time.

\bibliographystyle{IEEEtran}
\bibliography{Main}

\begin{thebibliography}{10}
\providecommand{\url}[1]{#1}
\csname url@samestyle\endcsname
\providecommand{\newblock}{\relax}
\providecommand{\bibinfo}[2]{#2}
\providecommand{\BIBentrySTDinterwordspacing}{\spaceskip=0pt\relax}
\providecommand{\BIBentryALTinterwordstretchfactor}{4}
\providecommand{\BIBentryALTinterwordspacing}{\spaceskip=\fontdimen2\font plus
\BIBentryALTinterwordstretchfactor\fontdimen3\font minus \fontdimen4\font\relax}
\providecommand{\BIBforeignlanguage}[2]{{%
\expandafter\ifx\csname l@#1\endcsname\relax
\typeout{** WARNING: IEEEtran.bst: No hyphenation pattern has been}%
\typeout{** loaded for the language `#1'. Using the pattern for}%
\typeout{** the default language instead.}%
\else
\language=\csname l@#1\endcsname
\fi
#2}}
\providecommand{\BIBdecl}{\relax}
\BIBdecl

\bibitem{gandomi3d}
K.~Gandomi, Z.~Zhao, M.~Tarasek, E.~Fiveland, C.~Bhushan, G.~Ghoshal, P.~Neubauer, E.~Williams, M.~Liu, P.~Carvalho \emph{et~al.}, ``3d thermo-acoustic modeling of a piezoelectric transducer for directed interstitial ultrasound ablation.''

\bibitem{rao2014magnetic}
M.~S. Rao, E.~L. Hargreaves, A.~J. Khan, B.~G. Haffty, and S.~F. Danish, ``Magnetic resonance-guided laser ablation improves local control for postradiosurgery recurrence and/or radiation necrosis,'' \emph{Neurosurgery}, vol.~74, no.~6, pp. 658--667, 2014.

\bibitem{carpentier2011laser}
A.~Carpentier, R.~J. McNichols, R.~J. Stafford, J.-P. Guichard, D.~Reizine, S.~Delaloge, E.~Vicaut, D.~Payen, A.~Gowda, and B.~George, ``Laser thermal therapy: Real-time mri-guided and computer-controlled procedures for metastatic brain tumors,'' \emph{Lasers in surgery and medicine}, vol.~43, no.~10, pp. 943--950, 2011.

\bibitem{ghoshal2013ex}
G.~Ghoshal, V.~Salgaonkar, J.~Wooton, E.~Williams, P.~Neubauer, L.~Frith, B.~Komadina, C.~Diederich, and E.~C. Burdette, ``Ex-vivo and simulation comparison of multi-angular ablation patterns using catheter-based ultrasound transducers,'' in \emph{Energy-based Treatment of Tissue and Assessment VII}, vol. 8584.\hskip 1em plus 0.5em minus 0.4em\relax SPIE, 2013, pp. 293--303.

\bibitem{missios2015renaissance}
S.~Missios, K.~Bekelis, and G.~H. Barnett, ``Renaissance of laser interstitial thermal ablation,'' \emph{Neurosurgical focus}, vol.~38, no.~3, p. E13, 2015.

\bibitem{christian2014focused}
E.~Christian, C.~Yu, and M.~L. Apuzzo, ``Focused ultrasound: relevant history and prospects for the addition of mechanical energy to the neurosurgical armamentarium,'' \emph{World Neurosurgery}, vol.~82, no. 3-4, pp. 354--365, 2014.

\bibitem{gandomi2019thermo}
K.~Gandomi, P.~Carvalho, Z.~Zhao, C.~Nycz, E.~Burdette, and G.~Fischer, ``Thermo-acoustic simulation of a piezoelectric transducer for interstitial thermal ablation with mrti based validation,'' in \emph{Comsol Conference}, 2019.

\bibitem{gandomi2020modeling}
K.~Y. Gandomi, P.~A. Carvalho, M.~Tarasek, E.~W. Fiveland, C.~Bhushan, E.~Williams, P.~Neubauer, Z.~Zhao, J.~Pilitsis, D.~Yeo \emph{et~al.}, ``Modeling of interstitial ultrasound ablation for continuous applicator rotation with mr validation,'' \emph{IEEE Transactions on Biomedical Engineering}, vol.~68, no.~6, pp. 1838--1846, 2020.

\bibitem{zhao2024deep}
Z.~Zhao, B.~Szewczyk, M.~Tarasek, C.~Bales, Y.~Wang, M.~Liu, Y.~Jiang, C.~Bhushan, E.~Fiveland, Z.~Campwala \emph{et~al.}, ``Deep brain ultrasound ablation thermal dose modeling with in vivo experimental validation,'' \emph{arXiv preprint arXiv:2409.02395}, 2024.

\bibitem{szewczyk2022happens}
B.~Szewczyk, M.~Tarasek, Z.~Campwala, R.~Trowbridge, Z.~Zhao, P.~M. Johansen, Z.~Olmsted, C.~Bhushan, E.~Fiveland, G.~Ghoshal \emph{et~al.}, ``What happens to brain outside the thermal ablation zones? an assessment of needle-based therapeutic ultrasound in survival swine,'' \emph{International Journal of Hyperthermia}, vol.~39, no.~1, pp. 1283--1293, 2022.

\bibitem{sadeghi2016parameter}
M.~Sadeghi-Goughari, A.~Mojra, and S.~Sadeghi, ``Parameter estimation of brain tumors using intraoperative thermal imaging based on artificial tactile sensing in conjunction with artificial neural network,'' \emph{Journal of Physics D: Applied Physics}, vol.~49, no.~7, p. 075404, 2016.

\bibitem{de2016heat}
M.~M. de~Oliveira, P.~Wen, and T.~Ahfock, ``Heat transfer due to electroconvulsive therapy: Influence of anisotropic thermal and electrical skull conductivity,'' \emph{Computer methods and programs in biomedicine}, vol. 133, pp. 71--81, 2016.

\bibitem{vaupel1989blood}
P.~Vaupel, F.~Kallinowski, and P.~Okunieff, ``Blood flow, oxygen and nutrient supply, and metabolic microenvironment of human tumors: a review,'' \emph{Cancer research}, vol.~49, no.~23, pp. 6449--6465, 1989.

\bibitem{bousselham2018brain}
A.~Bousselham, O.~Bouattane, M.~Youssfi, and A.~Raihani, ``Brain tumor temperature effect extraction from mri imaging using bioheat equation,'' \emph{Procedia Computer Science}, vol. 127, pp. 336--343, 2018.

\bibitem{tanaka1965localization}
K.~Tanaka, K.~Ito, and T.~Wagai, ``The localization of brain tumors by ultrasonic techniques: A clinical review of 111 cases,'' \emph{Journal of Neurosurgery}, vol.~23, no.~2, pp. 135--147, 1965.

\bibitem{campwala2021predicting}
Z.~Campwala, B.~Szewczyk, T.~Maietta, R.~Trowbridge, M.~Tarasek, C.~Bhushan, E.~Fiveland, G.~Ghoshal, T.~Heffter, K.~Gandomi \emph{et~al.}, ``Predicting ablation zones with multislice volumetric 2-d magnetic resonance thermal imaging,'' \emph{International Journal of Hyperthermia}, vol.~38, no.~1, pp. 907--915, 2021.

\end{thebibliography}

\end{document}